\documentclass[sigconf]{acmart} 

\usepackage[ruled,vlined,linesnumbered]{algorithm2e}

\AtBeginDocument{%
  \providecommand\BibTeX{{%
    \normalfont B\kern-0.5em{\scshape i\kern-0.25em b}\kern-0.8em\TeX}}}

\copyrightyear{2021}
\acmYear{2021}
\setcopyright{rightsretained}
\acmConference[CHI '21]{CHI Conference on Human Factors in Computing Systems}{May 8--13, 2021}{Yokohama, Japan}
\acmBooktitle{CHI Conference on Human Factors in Computing Systems (CHI '21), May 8--13, 2021, Yokohama, Japan}\acmDOI{10.1145/3411764.3445043}
\acmISBN{978-1-4503-8096-6/21/05}

\begin{document}

\title{ReverseORC: Reverse Engineering of Resizable User Interface Layouts with OR-Constraints}


\author{Yue Jiang}
\affiliation{%
  \institution{Max Planck Institute for Informatics}
  \city{Saarbrücken}
  \country{Germany}}
\email{yuejiang@mpi-inf.mpg.de}

\author{Wolfgang Stuerzlinger}
\affiliation{%
  \institution{School of Interactive Arts + Technology (SIAT), Simon Fraser University}
  \city{Vancouver}
  \country{Canada}}
\email{w.s@sfu.ca}

\author{Christof Lutteroth}
\affiliation{%
  \institution{Department of Computer Science, University of Bath}
  \city{Bath}
  \country{UK}}
\email{c.lutteroth@bath.ac.uk}

\renewcommand{\shortauthors}{Yue Jiang, Wolfgang Stuerzlinger, Christof Lutteroth}

\begin{abstract}
Reverse engineering (RE) of user interfaces (UIs) plays an important role in software evolution. However, the large diversity of UI technologies and the need for UIs to be resizable make this challenging. 
We propose ReverseORC, a novel RE approach able to discover diverse layout types and their dynamic resizing behaviours independently of their implementation, and to specify them by using OR constraints. Unlike previous RE approaches, ReverseORC infers flexible layout constraint specifications by sampling UIs at different sizes and analyzing the differences between them. It can create specifications that replicate even some non-standard layout managers with complex dynamic layout behaviours. We demonstrate that ReverseORC works across different platforms with very different layout approaches, {\em e.g.}, for GUIs as well as for the Web. Furthermore, it can be used to detect and fix problems in legacy UIs, extend UIs with enhanced layout behaviours, and support the creation of flexible UI layouts. 
\end{abstract}


\newcommand\DELETE[1]{\textcolor{red}{#1}}
\newcommand\ADD[1]{\textcolor{black}{#1}}
\newenvironment{delete}{\par\color{red}}{\par}
\newenvironment{add}{\par\color{black}}{\par}



\begin{teaserfigure}
  \includegraphics[width=\textwidth]{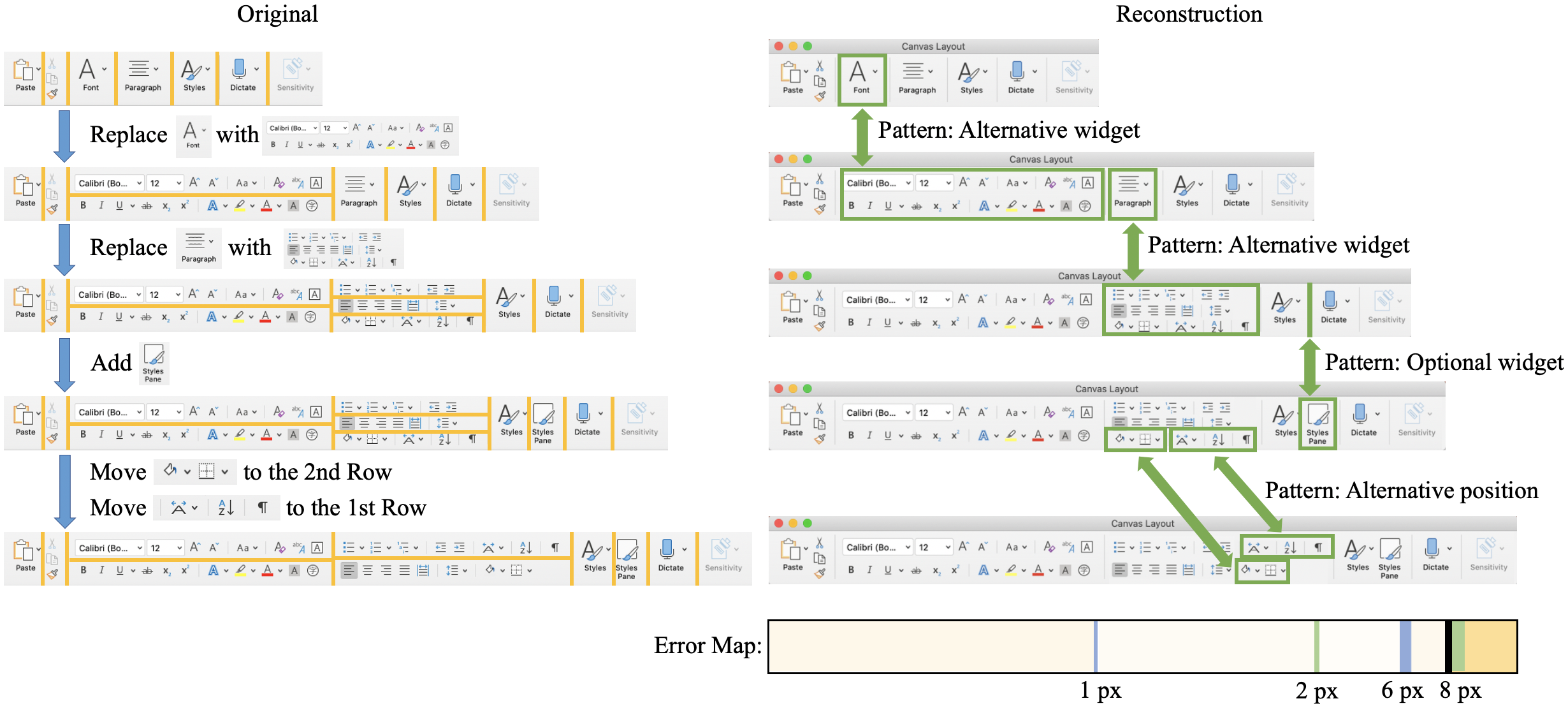}
  \caption{Reverse engineering of the MS Word ``ribbon'' toolbar. ReverseORC samples the user interface (UI) at different sizes and reconstructs parsimonious layout specifications for each size. It then detects changes between the layout specifications using a novel diff algorithm for layouts, and matches the changes with corresponding layout patterns to reconstruct a UI with the same resize behaviours as the original. 
  We visualize the overall quality of reconstruction at different sizes in an error map by color-coding structural error (shades of yellow), transition error (blue/green), and `fault lines' (black) indicating potentially inconsistent behaviors.
  }
  \Description{This figure shows the appearance of original the MS Word ``ribbon'' toolbar and the reconstructed MS Word ``ribbon'' toolbar using ReverseORC at five different sizes. Horizontal lines on the original toolbar are divisors that define Row layout containers and vertical ones define Columns. We show the changes between neighbouring sizes of the original toolbar and the corresponding layout patterns between neighbouring sizes of the reconstructed toolbar. An error map visualizes the overall quality of reconstruction including structural error (shades of yellow), transition error (blue/green), and `fault lines' (black) indicating potentially inconsistent behaviors}
  \label{fig:teaser}
\end{teaserfigure}

\newcommand\es[1]{\textcolor{blue}{#1}}

\maketitle

\section{Introduction}

Modern user interfaces (UIs) have become increasingly flexible. They use sophisticated layouts that can adapt to different sizes and orientations. For example, responsive web layouts \cite{marcotte2011responsive} enable designers to create web UIs that work on large desktop screens, small tablets, and tiny mobile devices by rearranging and adapting the UI. Similar approaches are used for mobile UIs \cite{sahami2013insights, zeidler2017automatic} and sometimes even for desktop UIs. The UIs and their layouts are created using UI toolkits~\cite{myers2000past, hudson1997supporting, hudson2000providing} and UI builders~\cite{zeidler2012auckland, scoditti2009new}, 
which facilitate the efficient creation and editing of common layouts and support an iterative design process. However, available UI toolkits, builders and supported layouts are numerous and constantly evolving, leading to a wide diversity of different layouts such as grid, flow, group, stack, tile, flexbox and constraint-based layouts. 

It is challenging to change an existing UI if its source code or specification is not available. Even if a specification is available, it is usually tied to the UI toolkit that was used to create the UI, and the diversity of UI toolkits and layouts makes it hard to understand and use such a specification. The problem of reconstructing an existing UI for further development is called UI {\it reverse engineering} (RE). It is known to be difficult but often necessary as software and devices evolve and new UI toolkits and platforms need to be supported. For example, developers may want to modernize a legacy UI to benefit from novel technologies -- a use case where it is quite common that source code is not available, hard to modify, or an equivalent layout API may not exist. An illustrative case here is porting desktop GUIs to smartphones or vice versa, or even to/from webpages. It is very challenging to reverse engineer a UI so that its features and behaviours are consistent across different toolkits and platforms; so developers usually spend a lot of time trying to understand a specification and often end up manually re-implementing large portions of the UI~\cite{greenberg2008usability, hudson2005extensible, olsen2007evaluation}. 

In order to ease the burden of UI RE, many automated RE tools have been proposed. By using automated RE tools, specifications of GUI elements, layouts, and application behaviours can be extracted and modified independently of their underlying implementations and platforms. Previous works on reverse engineering UIs focus on detecting components in the interface~\cite{moore1996rule, Stuerzlinger2006facades,dixon2010prefab}, migrating UIs from one platform to another~\cite{moore1998user, moore1997using, moore2000miigrating, gerdes2009user, ramon2014model, ramon2016layout, christof2008automated}, and/or performing input and output redirection~\cite{Stuerzlinger2006facades, dixon2012general, dixon2014pixel, swearngin2017genie}. Previous works have shown that RE tools can reconstruct UI layouts that look similar to the originals, and can then generate implementations of the UI for other UI platforms and toolkits. However, while UI layout has evolved, RE tools have not kept pace with modern UIs: they cannot currently capture the complex resizing behaviours that have become commonplace for the web, on mobile devices, and even many desktop UIs. 

This paper presents ReverseORC, a novel prototype that is able to reverse engineer UI layout specifications based on a UI's resize behaviors. Given only widget placements for different UI sizes of an existing GUI, ReverseORC identifies how layout behaviours are encoded in the UI and generates a corresponding layout specification. The new layout specification is expressed using ORC Layout~\cite{jiang2019orclayout}, an approach for constraint-based layouts based on OR-constraints (ORC). ORC Layout is a powerful tool that allows designers to express modern resizable UI layouts in a mathematical, platform-independent manner, as constraint optimization problems. It unifies flow layouts and conventional constraint-based layouts to represent a large variety of layouts for desktop, web and mobile platforms. We build ReverseORC on ORC Layout as this is one of the most flexible layout specification mechanisms that does not involve programming.

ReverseORC automatically extracts specifications of how a UI is laid out for different UI sizes. It determines which UI size samples are necessary to infer an equivalent ORC layout specification. Although it is not possible in general to reverse engineer an arbitrary layout algorithm solely from examples for its output, ReverseORC is able to detect common layouts such as grids and flow. Additionally, it is able to detect advanced patterns describing dynamic changes in a layout (ORC patterns), such as widgets shifting or disappearing when the UI is made smaller. ReverseORC is designed to generate parsimonious specifications, i.e.\ specifications that are sufficiently rich to capture the desired behaviour, but not more complex than necessary. This makes ReverseORC's output easier to understand for a human designer, so that they can potentially build on it later. Only if ReverseORC cannot identify a pattern in the observed changes, then it uses OR constraints to combine the specifications of the respective layouts. 

We demonstrate that ReverseORC can be applied to UIs on different platforms, such as desktop and web UIs, reconstructing platform-independent specifications for a wide range of UI technologies. Furthermore, we support a variety of use cases based on the generated ORC layout specifications: Many existing GUI layouts are static or cannot fit a large range of screen sizes adaptively, {\em e.g.}, from smart watches and smartphones to desktop environments. Designers can specify desired adaptations of such a layout by example, and let ReverseORC reconstruct an appropriate flexible layout specification. In a similar manner, designers can use ReverseORC to create new resizable layout specifications from scratch. Furthermore, ReverseORC allows designers to fix bad layout behaviors by modifying the generated ORC layout specification. In summary, ReverseORC lifts the level of abstraction of the layout specification process by allows designers to create and edit sophisticated flexible layout specifications by example. 

{\it Novelty}. In contrast to existing GUI reverse engineering approaches~\cite{moore1998user, moore1997using, moore2000miigrating, gerdes2009user, ramon2014model, ramon2016layout, christof2008automated}, ReverseORC is able to reconstruct the {\it dynamic} resizing behaviors of a GUI. Extending previous work that identified static layout components, it detects advanced layout patterns such as optional and shifting widgets and specifies their behaviors. It is platform and toolkit independent, which enables reuse of layouts across applications and platforms. Finally, it allows designers to specify the resize behavior of UIs by example. 
In particular, we demonstrate the following contributions: 
\begin{enumerate}
\item A novel method for identifying and reverse engineering dynamic layout behaviors for different platforms, only by sampling different layout sizes for an existing UI.
\item A novel method for detecting the differences between layout specifications.
\item A novel method of specifying and editing resizable layouts by example.
\item Validation of our approach based on real-life layouts, including GUI layouts, {\em e.g.}, the Microsoft Word Ribbon, and web layouts, {\em e.g.}, the BBC News website, as well as demonstrating that ReverseORC can reverse engineer layouts based on a very small number of exemplars. 
\end{enumerate}

\section{Related Work}


\subsection{Resizable UI Layout}

Due to the large diversity of existing computing devices, which vary in their screen sizes and aspect ratios, and users' different personal viewing preferences, it is important that applications support resizable UI layouts. Layout models are widely used to specify resizable UI layouts, and layout managers then generate the layout results based on the specifications. Early approaches proposed simple layout models, such as group, grid, table, and grid-bag layouts~\cite{myers2000past, myers1995user}. Object-oriented models like Amulet~\cite{myers97theamulet} combined properties of flow and grid layouts. Modern GUI layout models are mostly constraint-based~\cite{zeidler2017tiling, lutteroth2008domain} and used together with UI builders, which can create layout constraints based on direct manipulation~\cite{karsenty1993inferring, scoditti2009new, weber2010reduction, zeidler2012auckland}.

Jiang et al.~\cite{jiang2019orclayout} proposed ORC Layout, an approach for constraint-based layouts based on OR-constraints (ORC). An OR-constraint is a disjunctive constraint, where only one disjunctive part needs to be true. \ADD{ORC Layout unifies flow and conventional constraint-based layouts through adding OR-constraints to standard soft/hard linear constraint systems. ORC Layout specifications also enable the use of ORC design patterns, which enable designers to create a large variety of flexible layouts that work across different screen sizes and orientations.} ORC Layout is a powerful, high-level layout specification method; it enables users to describe layouts with dynamic behaviors that adapt to screens with very different sizes, orientations, and aspect ratios, using only a single layout specification.
ORCSolver~\cite{jiang2020orcsolver} is a novel solving technique to efficiently solve ORC Layout specifications. 
\ADD{ReverseORC uses ORC Layout to express the captured dynamic layout behaviors.} 

Previous work also investigated resizable web layouts. Chen et al.~\cite{chen2005adapting} presented a page-adaptation technique that splits a web page into smaller blocks to adapt pages for small screen devices. Xie et al.~\cite{xie2005adaptive} proposed a novel document representation dynamically adapting screen sizes. Domshlak~\cite{domshlak2000preference} enabled personalized presentation of web page content. Constraints can also be used to specify the desired layout of a web page, e.g.\ Borning et al.~\cite{borning2000constraint} proposed a constraint-based web system enabling both the author and the viewer to define page layout constraints. Hosobe~\cite{hosobe2005solving} introduced an algorithm to solve hybrid systems of linear constraints and one-way constraints to handle web document layouts efficiently.

\subsection{Customizing User Interfaces}

Researchers have proposed several approaches that can be used to modify a GUI if it is not automatically adapted to the user's requirements or if the adaptation is sub-optimal, {\em e.g.,} when using a GUI on a device with a smaller size.
For traditional GUIs, Edwards et al.~\cite{edwards1997systematic} and Olsen et al.~\cite{olsen1999implementing} proposed to modify interfaces by replacing drawing objects and intercepting API commands in applications with specific toolkit implementations. WinCuts~\cite{tan2004wincuts} enabled window subdivision with a copy-paste method to configure input/output redirection. Mudibo~\cite{hutchings2005mudibo} used input/output redirection to generate windows with multiple alternative positions, and allowed users to choose a desired one. User Interface Fa\c{c}ades~\cite{Stuerzlinger2006facades} detected all widgets and their hierarchy through an accessibility API, enabled widget replacement, and presented advanced customization of runtime interaction behaviour.

Previous research on web UI customization was mostly based on a structured presentation, the Document Object Model (DOM). ChickenFoot~\cite{bolin2005automation}, CoScripter~\cite{leshed2008CoScripter}, and Koala~\cite{little2007koala} automated, customized, and integrated web applications. Clip, Connect, Clone~\cite{fujima2004clip}, d.mix~\cite{hartmann2007programming}, and Vegemite~\cite{lin2009end} introduced end-user mash-up methods between existing applications. Highlight~\cite{nichols2008mobilization} re-authored web applications on mobile interfaces. 

As all UIs are instantiated as pixels, previous work widely explored pixel-level interpretation to enhance UIs. Pixel-based approaches have been proposed to access data~\cite{potter1992triggers}, record the actions performed by users~\cite{st2000programming}, translate input and output into different forms~\cite{amant2005image}, improve target detection in accessibility APIs~\cite{hurst2010automatically}, perform visual manipulation~\cite{zettlemoyer1999visual}, and event management~\cite{zettlemoyer1998ibots}. ScreenCrayons~\cite{olsen2004screencrayons} enabled document and visual annotation. Sikuli~\cite{chang2010gui, yeh2009sikuli} supported UI testing by writing visual test scripts. Genie~\cite{swearngin2017genie} reverse engineered underlying commands to enable users to engage with web applications via different input modalities. Hurst et al.~\cite{hurst2010automatically} presented improved target boundary detection based on the combination of an accessibility API and pixel-based methods.
\ADD{ReverseORC is a platform- and framework-independent system enabling customization for both GUI layouts and web layouts.}

\subsection{Reverse Engineering}

\begin{table}[t]
\caption{\ADD{Overview of reverse engineering approaches (`+' denotes full and `$\sim$' partial support for certain layouts).}}
\begin{add}
  \centering
 \begin{tabular}{ | l | c | c | c | }
    \hline
     & Basic & Flow & Dynamic \\
     & Layouts & Layouts & Topology
     \\ \hline
     Most previous RE work & + & &
     \\ \hline
Constraint-based RE~\cite{christof2008automated}   & + & &
\\ \hline

  Model-driven RE~\cite{ramon2014model, ramon2016layout} & + & $\sim$ &
  \\ \hline
  Expresso~\cite{krosnick2018expresso}  & & & $\sim$
\\ \hline
 InferUI~{\cite{bielik2018robust}} & + & & 
 
\\ \hline

\textbf{ReverseORC} & + & + & +
  
\\ \hline

  \end{tabular}
  \label{tbl:comparison}
  \end{add}
\end{table}

UI reverse engineering is widely used to migrate applications from one platform to another. Moore~\cite{moore1996rule} presented a rule-based detection approach for partially automating the process of reversing engineering legacy applications. Staiger~\cite{staiger2007static} analyzed the source code, identified widgets, and reconstructed the GUI tree. MORPH~\cite{moore1998user, moore1997using} proposed a model-oriented re-engineering process for migrating character-based legacy UIs to GUIs. REMAUI~\cite{nguyen2015reverse} was a pixel-based approach that automatically reverse engineered mobile application UIs. None of the above approaches yielded resizable layout information.

Reverse engineering has been used as a way to perform GUI customization. UI Fa\c{c}ades~\cite{Stuerzlinger2006facades} enabled users to replace widgets and change application behaviors for an existing application at runtime through an accessibility API-based approach. Prefab~\cite{dixon2010prefab, dixon2012general, dixon2014pixel, dixon2011content, dixon2014prefab} was a pixel-based approach that provided a tree structure to interpret content and hierarchy~\cite{dixon2011content}. Both approaches identify interface elements and allow the user to add interactive enhancements to a GUI~\cite{Stuerzlinger2006facades,dixon2010prefab, dixon2014prefab}. However, none of these approaches allowed users to modify the layout itself. 
Instead of using pixel-based interpretations of a UI for reverse engineering~\cite{dixon2014prefab} or migrating a UI directly between different platforms~\cite{gerdes2009user, sanchez2010model}, our approach detects layout behaviors and generates standard ORC Layout specifications to facilitate UI development and customization. 
\ADD{Similar to ReverseORC’s layout structure reconstruction, InferUI~{\cite{bielik2018robust}} infers constraints to describe a layout from UI exemplars. Yet, InferUI generates only linear constraints, which maintain relative mutual alignments of widgets but can only express a single topological arrangement. In contrast, ReverseORC infers OR-constraints, which can express dynamic topological layout changes such as flow, optional widgets, and alternative positions.}

Lutteroth~\cite{christof2008automated} reverse engineered GUI layouts to recover higher-level constraint-based specifications~\cite{lutteroth2006user} and to generate layouts that are resizable. S\'{a}nchez Ram\'{o}n et al.~\cite{sanchez2010model,ramon2014model, ramon2016layout} proposed a model-driven approach to reverse engineer legacy GUIs by capturing the visual arrangement of elements in the layout and produced GUI models with that explicit layout. 
\ADD{While these approaches~{\cite{ramon2014model, ramon2016layout}} can capture common layout containers in a hierarchical manner, ReverseORC is also able to reconstruct a platform-independent specification of dynamic UI changes, such as optional widgets or widgets that change position across the layout hierarchy to accommodate changes in screen space, which cannot be expressed with common layout containers.}
The above approaches were only able to deal with simple layout behaviors such as grid arrangements, but could not deal with layouts that included dynamic layout changes such as flows, shifting widgets or optional widgets. 

Reverse engineering is also useful for web layouts. Moore et al.~\cite{moore2000miigrating} used the MORPH technique~\cite{moore1998user, moore1997using} to re-engineer legacy information systems to operate on the web. CELLEST~\cite{stroulia2003user} demonstrated a process for migrating legacy GUIs to web-accessible platforms. Gerdes~\cite{gerdes2009user} proposed a method to migrate Windows applications to Visual Basic .NET, based on runtime traces. VAQUISTA~\cite{vanderdonckt2001flexible} reverse engineered the presentation model of a web page to generate equivalent GUIs for other platforms. VIPS~\cite{cai2003vips} presented an approach for web content structure analysis based on visual representation. 
\ADD{Similar to ReverseORC’s exemplar-based layout design, Expresso~{\cite{krosnick2018expresso}} allows designers to specify samples of a web UI at different sizes. Expresso then either linearly interpolates widget positions and sizes between the given UI sizes (‘keyframes’), or lets them jump discontinuously, as specified by the designer. In contrast to ReverseORC, Expresso does not infer behavioral UI layout patterns dynamically. For example, if widgets should flow onto a new line, the designer would have to specify keyframes for every possible line break in Expresso.}
To the best of our knowledge there is no reverse engineering approach for UI layouts that can extract a UI's dynamic resize behaviours. \ADD{\autoref{tbl:comparison} shows a comparison of the capabilities of different reverse engineering approaches.}

\begin{add}

\begin{figure*}[t]
\centering
\includegraphics[width=\textwidth]{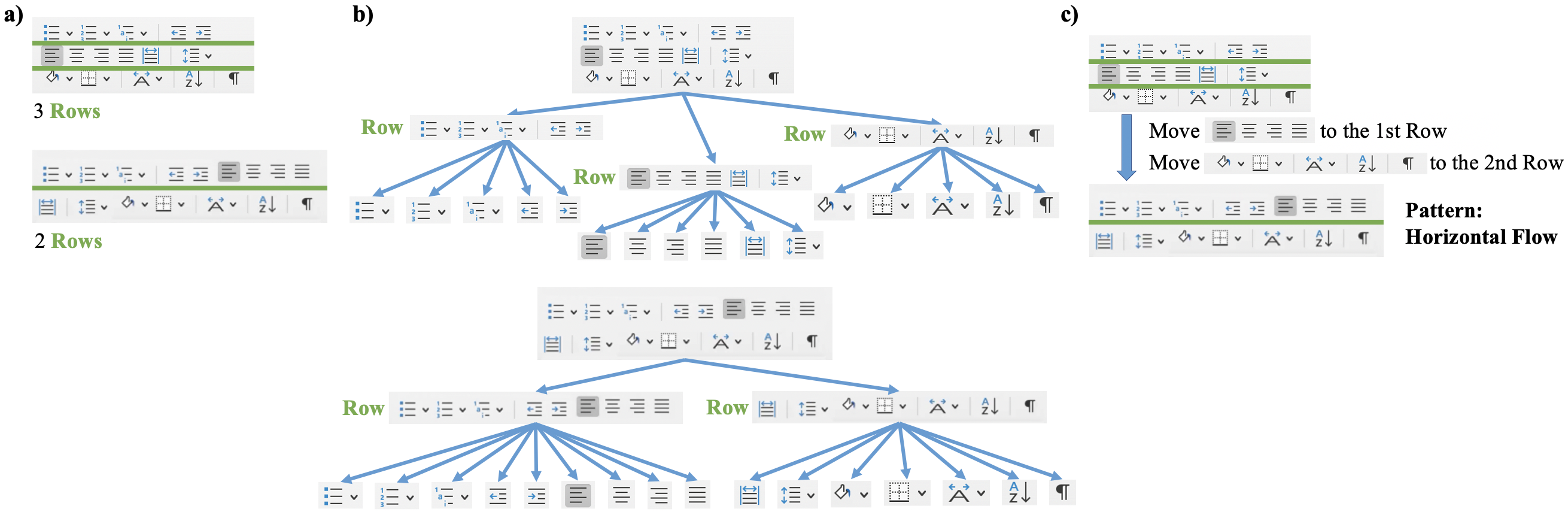}
\caption{\ADD{Overview of the different stages of the ReverseORC approach: a) UI sampling by setting different window sizes (horizontal lines are divisors that define {\it Row} layout containers); b) layout tree structure construction for each sample; and c) difference detection between layout trees and layout pattern inference.}}

\Description{This figure shows a example that illustrates each stage of the process. a) shows two different sampling results at different window sizes. The first exemplar shows a 3-row toolbar and the second exemplar shows the same toolbar with two rows at a smaller window height. b) shows the corresponding layout trees constructed corresponding to each layout exemplar. c) shows the differences between the two layout trees and refers that the layout behaviours match a horizontal flow layout pattern.}
\label{fig:overview}
\end{figure*}

\section{Overview}

Our ReverseORC approach first extracts widget information from the layout through accessibility APIs. 
Then, it uses a grid search to sample and resize the layout through setting different window sizes (\autoref{fig:overview} a). It constructs a layout tree for each sampled layout (\autoref{fig:overview} b). ReverseORC then tracks all differences between layout trees of neighboring layouts during the sampling process and generates corresponding layout differences. Based on these layout differences, ReverseORC then infers overall layout behaviors and patterns, and constructs a corresponding ORC layout specification, enabling later modification and customization (\autoref{fig:overview} c).

\subsection{Usage Scenarios}

ReverseORC fits into standard software development practice and has many practical applications. Some typical usage scenarios are:

\begin{enumerate}
    \item Developers initiate automatic UI sampling with a platform-specific tool: For desktop UIs, developers run the app to reverse engineer. Our tool then uses OS calls to set the UI window size and accessibility API calls to query widgets and their properties. For web UIs, developers use a tool with an embedded browser, instrumented to resize and extract widgets after the UI has been loaded. Similarly, for mobile UIs we use an emulator, with emulator calls to set the screen size and extract widgets. For each UI sample, all widgets and their properties are stored in a platform independent format. Previous work has demonstrated such approaches.
    
    \item Layout structure reconstruction, difference detection, and ORC specification generation are performed automatically based on the UI samples with our platform-independent tool. The tool visualises the quality of the reverse engineered UI (see 6.2) and allows developers to display it at specific sizes by clicking on points on the error map. Developers can adjust the results and fix bad layout behaviors by modifying and adding UI samples (see 7.3), or editing the ORC specification directly using the ORC Editor~{\cite{jiang2019orclayout, jiang2020orcsolver}}.
    
    \item ORC UI specifications can then be used directly by running them on a platform-specific implementation of the ORC Solver, which can run on desktop and mobile platforms as well as the web{\footnote{github.com/cpitclaudel/z3.wasm}}.
\end{enumerate}

 \end{add}



\section{User Interface Sampling}  

To reverse engineer a GUI, we follow the common approach of first detecting the widgets of the UI, and then reconstructing the layout of the widgets using ORC Layout as abstraction model. Subsequently, we transform the reconstructed specification to its target form, generating a new UI for the desired platform. As discussed above, previous works only reconstruct lower-level UI specifications that ignore the more abstract aspects of UI layout during this process.
By sampling an existing UI at different layout sizes, ReverseORC can identify and reverse engineer both GUI layout and web layout behaviours. It first extracts widget information from the layout through accessibility APIs. Then, it uses a grid search to sample and resize the UI layout by setting different window or screen sizes. ReverseORC keeps track of any differences between neighbouring layouts during the sampling process, and is then able to reconstruct an abstract layout specification based on the way the layout changes depending on its size.

\subsection{Widget Extraction}

Similar to UI Fa\c{c}ades~\cite{Stuerzlinger2006facades}, our approach extracts widget information of UIs through an accessibility API. An accessibility API provides a structured representation analogous to the Document Object Model (DOM). Compared to pixel-based approaches~\cite{dixon2010prefab} or computer vision recognition-based approaches~\cite{nguyen2015reverse}, accessibility APIs directly access the underlying data of a UI, which avoids the potential for recognition errors. In addition, accessibility APIs can access information that is not visible or not obtainable by analyzing raw pixels, {\em e.g.}, widget identities. We still acknowledge that pixel-based approaches could be used as an alternative mechanism in ReverseORC, albeit at the price of an increased risk of layout recognition errors.

To extract the widget information ReverseORC needs, we traverse the structured representation through the accessibility API. Under the assumption that the bounding box of each widget is rectangular, for each widget $w$ in the layout, we retrieve its unique identifier ($w.id$), size ($w.\textit{width}$, $w.\textit{height}$), and coordinates for its top-left corner ($w.\textit{left}$, $w.\textit{top}$). 
Some accessibility APIs provide more information about a UI, including information not only about the widgets but also about the layout managers used. For example, it is generally possible to access the full DOM of a web UI. However, ReverseORC does not use this information for the following reasons: 1) Layout information is not always available, {\em e.g.}, some desktop UIs do not provide it. 2) There are too many layout managers to understand the layout behavior of a UI just from the DOM, so DOM layout containers are often like black boxes. 3) Layout behaviors are often described at least partly programmatically rather than in the DOM, {\em e.g.}, using JavaScript code; therefore they cannot be inferred from the DOM alone. And 4) even if we could interpret a DOM description of a UI layout, DOMs are often much more complicated than they need to be. For example, many complex web apps use large numbers of nested DIV elements, confounding aspects of layout and functional application design. One of the aims of ReverseORC is to provide a parsimonious layout representation, {\em i.e.}, a representation that avoids unnecessary complexities. This is achieved by analysing not how developers have specified layouts, but by analysing what layouts actually looks like, in the simplest terms possible.

\subsection{Grid Search}

We use an adaptive grid search approach to obtain a representative set of different layout exemplars by resizing the window or a virtual screen. 
A brute force method to thoroughly analyze a layout would be to sample as many exemplars as possible. However, in practice, it can be expensive to resize the layout to all potential sizes, and it would create unnecessary work for the later reverse engineering stages. Thus, it is best to minimize the number of queries by taking advantages of the continuous nature of UI layout: layout changes occur incrementally, as it would otherwise confuse the user. If two sampled layout exemplars have the same structure or their variance matches layout behaviours we have already detected, then there is no need to subsample further and to explore more exemplars in the range between the sizes of these two layout exemplars. In this case we (very likely) have already identified all the behaviors in this range and are unlikely to get more information by further subsampling. 

We chose to perform an adaptive grid search to sample UI layout, as layout sampling is a two-dimensional problem. Both the width and height of a UI are likely to affect its layout, with UIs often assuming different layouts for different sizes and aspect ratios.
We start with the extreme window sizes (minimum and maximum) and sample new layout sizes based on a binary grid search. We define the maximum size of a layout as the biggest screen size we would like to support, and the minimum size as is defined by the layout, i.e., the minimum size that the UI can be set to.
During the search process, if two sampled layout exemplars $L1$ and $L2$ have the same structure or their variance matches layout behaviors we have already detected, then we stop subsampling in the window size range between these two layout exemplars. Otherwise, we subsample depending on their sizes. If they have the same width but different heights, we subsample a layout exemplar with the same width and the middle height of the two. Analogously, if they have the same height but different widths, we subsample a layout exemplar with the same height and the middle width of the two. If both width and height are different, we subsample three layout exemplar with 1) middle width and middle height, 2) same width as $L1$ and same height as $L2$, and 3) same height as $L1$ and same width as $L2$, respectively.




\begin{figure*}[t]
\centering
\includegraphics[width=0.75\textwidth]{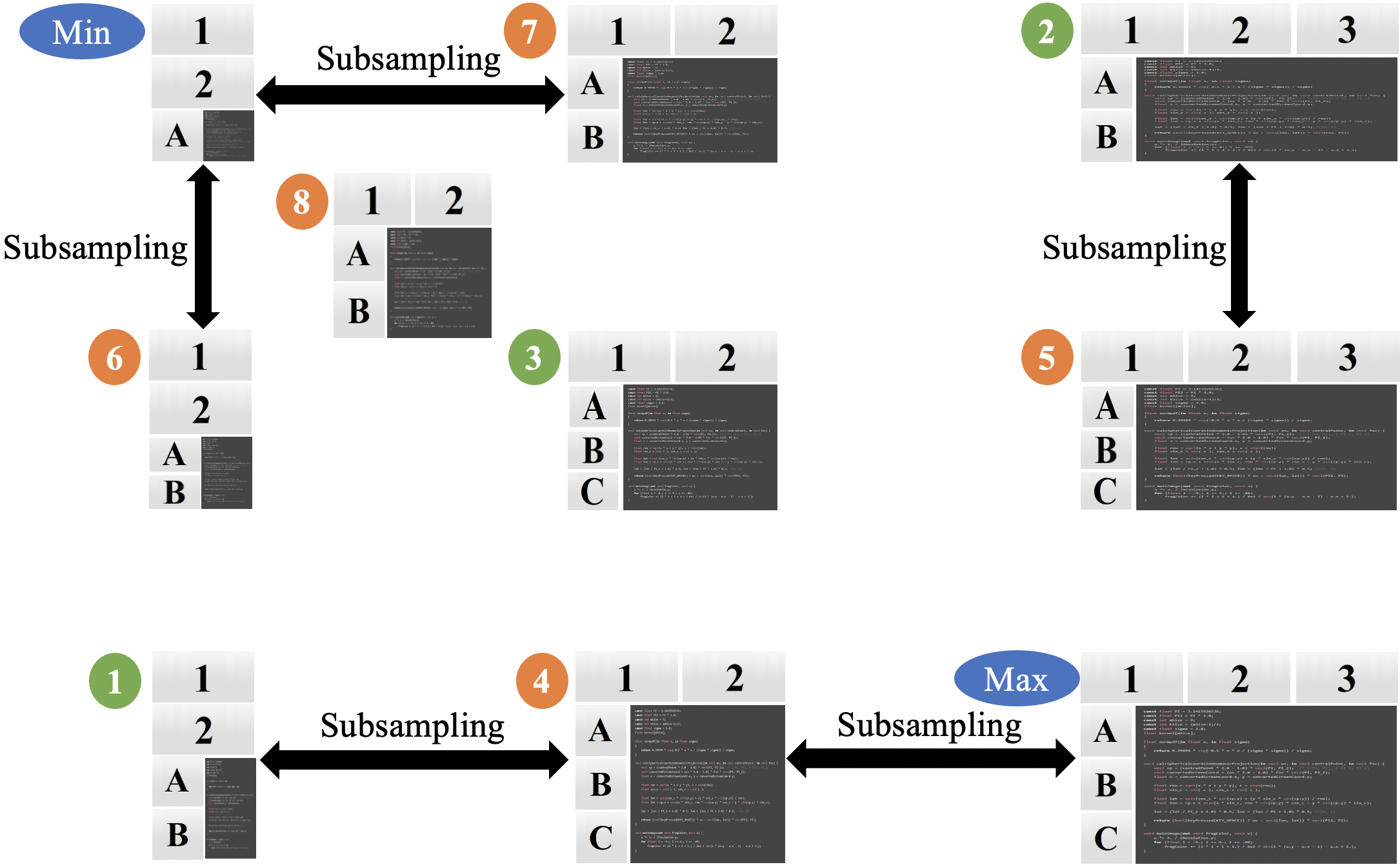}
\caption{\ADD{Grid subsampling example.}}

\Description{This figure shows a grid subsampling example. The minimum window size is at the left top corner, the maximum size is at the right bottom corner, and the subsampled window sizes are shown in between.}
\label{fig:sampling}
\end{figure*}

We show a sampling example in Figure \ref{fig:sampling}, referring to the layout exemplars as (Min), (Max), (1), (2), etc. We start with the extreme layout sizes, {\em i.e.}, the minimum size (Min) shown at the top left and the maximum (Max) at the bottom right. According to the subsampling rules, based on (Min) and (Max), we subsample layout exemplars (1), (2) and (3). To minimize the number of subsampling exemplars, we first subsample between two exemplars with the same height or width, and perform further subsampling along the diagonals between exemplars only if both variations in height and width show changes in the layout structures. For example, we subsample layout (4) between (Max) and (1). Widget 3 disappears in (4), so we continue to subsample between (Max) and (4) to detect the point of its disappearance, stopping the subsampling once the size difference between two exemplars is small. The structural difference between (Max) and (3) is the same as the difference between (Max) and (4), {\em i.e.}, widget 3 disappears. As we have already subsampled between (Max) and (4), we do not subsample further between (Max) and (3). We keep subsampling until we find all structural differences of the exemplars and the approximate transition points of changes.

\section{Layout Structure Reconstruction}

In order to compare the differences between layouts, we aim to reconstruct the simplest possible specification for the structure of a layout exemplar. We use symbolic tabstop dividers to divide layouts into separated parts in order to define layout structure. Such layout structure makes it easier to compare layouts and detect the differences between layouts.

\subsection{Tabstops}

A tabstop is an abstraction that has been introduced in previous work on GUI layouts~\cite{hashimoto1992graphical, hudson1990interactive, lutteroth2006user, zeidler2012auckland}. It is a symbolic object in the layout used to represent the alignments of multiple widgets. Associated with the two dimensions of the plane, there are two types of tabstops: x-tabstop and y-tabstop. An x-tabstop represents a position on the x-axis and correspondingly for a y-tabstop. Tabstops are in effect variables defining horizontal (y-tabstops) or vertical (x-tabstops) grid lines. The combination of x-tabstops and y-tabstops in a GUI forms a grid controlling how widgets are aligned in the GUI. Each widget $w$ has four tabstop variables ($w.\textit{left}$, $w.\textit{right}$, $w.\textit{top}$, $w.\textit{bottom}$) that delimit the area it occupies. Similarly, a layout $L$ itself has four tabstop variables $L.\textit{left}$, $L.\textit{right}$, $L.\textit{top}$ and $L.\textit{bottom}$ that define its boundaries, which is typically called the window (or panel) size.

The main advantage of using tabstops is that in a constraint-based layout system, if some widgets share a boundary, can just add a tabstop to the specification and then have all the corresponding widgets refer to that tabstop instead of adding separate alignment constraints for each widget. This approach makes it easier to maintain and modify the resulting constraint system. Whenever we need to change the alignment of the widgets sharing a tabstop, we just need to change the constraints relating to that tabstop, and then all the corresponding widgets will be positioned accordingly.
For each layout $L$, we define tabstops through two functions $xtabs()$ and $ytabs()$ that map from positions in the GUI to tabstop variables in the layout: $xtabs$ is a function mapping from x-coordinates to x-tabstops, and $ytabs$ mapps y-coordinates to y-tabstops (See Appendix~\ref{tabstopCreation} for details about tabstop creation).

To guide layout reconstruction, we call a tabstop a {\it layout divider} if it is a clean cut dividing the layout into two parts without crossing any widget in the layout. To reconstruct the containment hierarchy of a UI, the concept of layout dividers is applied recursively on the sublayouts contained in a layout. For example, in Figure~\ref{fig:treeStructure} the orange lines are the vertical layout dividers of the overall layout, and the green lines are horizontal layout dividers of sublayouts. 
Figure \ref{fig:tree} shows two examples of subdivision results. For a horizontal layout divider, all the widgets in the layout are either above it or below it, and analogously for vertical layout dividers (See Appendix~\ref{tabstopLayoutDivider} for details about tabstop layout divider detection).

\subsection{Reconstruction Algorithm}
\label{layoutStructureConstruction}

\begin{figure*}[t]
\centering
\includegraphics[width=\textwidth]{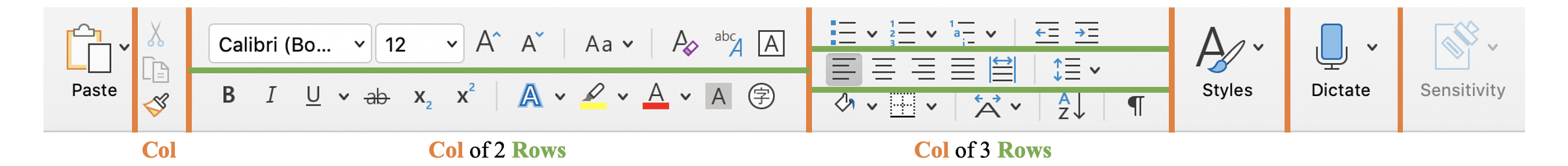}
\caption{Visualization of the layout structure of the MS Word ``ribbon'' as reconstructed in Section \ref{layoutStructureConstruction}. 
\ADD{Horizontal lines are divisors that define {\it Row} layout containers and vertical lines define {\it Col(umn)s}.}}

\Description{This figure shows a visualization of the layout structure of the MS Word ``ribbon''. On the layout, horizontal lines are divisors that define Row layout containers and vertical lines define Columns.}
\label{fig:treeStructure}
\end{figure*}

\begin{add}
Our layout structure reconstruction algorithm uses the same principles as the XY-Cut algorithm~\cite{nagy1984hierarchical, ha1995recursive} but works at a higher level of abstraction. Rather than segmenting an image based on gaps, we consider widget boundaries directly and we remove cuts if this allows us to simplify the X-Y structure.

We define layout structure using {\it Row} and {\it Column} layout containers. Two widgets belong to the same {\it Row} if they are located between the same two horizontal layout dividers, and analogously for {\it Column}. The resulting layout structure is a nested {\it Row} and {\it Column} structure.
We reconstruct the layout structure by recursively subdividing it based on layout dividers. 
We try horizontal subdivision (with vertical layout dividers) first as it is more common and in line with reading order. If horizontal subdivision is not possible, we process vertical subdivision analogously. 
We then assign the widgets to different sublayouts based on the positions of the horizontal layout dividers, and recursively use the reconstruction on each sublayout structure.
If both cases are impossible, which is very rare as UIs are typically laid out using a division-based containment hierarchy, then the layout can only be described using tabstops, {\em e.g.}, in a pinwheel layout \cite{zeidler2017tiling} (see Appendix~\ref{layooutStructureConstruction} for details about layout structure construction). 
Figure \ref{fig:treeStructure} shows the visualization of the reconstructed layout structure of the MS Word ``ribbon''. 

We aim to reconstruct the simplest possible layout structure. To avoid creating layout dividers caused by accidental alignments, we regroup widgets in multiple consecutive sublayouts and try running the algorithm recursively to simplify the resulting layout structure. We reconstruct the sublayout if we can get a simplified sublayout structure by grouping them.
\end{add}

\section{Layout Difference Detection}

ReverseORC keeps track of differences between neighbouring layouts during the sampling process and generates corresponding layout change sets. Based on the constructed layout structure, we can generate a corresponding layout specification tree where widgets are leaf nodes, and {\it Row}s and {\it Column}s are internal, non-leaf nodes. Our layout difference detection algorithm takes two layout trees as input and generates a set of edit operations that indicates the differences between the two. The set of edit operations is then used to infer layout behaviors.

Although there are many existing difference detection algorithms \cite{wang2003x, hashimoto2008diff, finis2013rws}, they cannot easily be applied because UI layouts have different requirements than other common tree structures such as XML or source code. Previous tree difference detection algorithms usually either consider none of the tree nodes to have a unique identity, or all of the nodes to have a unique identity. However, in the case of layouts, some tree nodes (the widgets) have a unique identity and some have not (the layout nodes). We can observe the widgets from the outside, {\em e.g.}, through an accessibility API, and can identify them. However, we cannot reliably identify layout elements such as rows and columns as accessibility APIs usually do not deliver this information; we infer their presence only by the way the UI is structured. So in a nutshell, our difference detection algorithm must be able to work with identities for some nodes, but not others.
In addition, previous difference detection algorithms often only supported the detection of deletion, insertion, and moving. For thoroughly analyzing and comparing layouts, we need more edit operations, such as whether a Row has changed to a Column. Furthermore, the computational complexity of generic tree difference detection algorithms are often quadratic. Our layout tree difference detection algorithm only takes linear time for detecting layout differences in practice.

\subsection{Edit Operations}

We define the following edit operations that can be applied to change a layout specification $S1$ to another specification $S2$, thus indicating the differences between them:
\begin{itemize}
    \item $\mathit{addNode}(s2)$: add node $s2$ in $S2$
    \item $\mathit{removeNode}(s1)$: remove node $s1$ from $S1$
    \item $\mathit{moveNode}(s1, s2)$: move node $s1$ in $S1$ to node $s2$ in $S2$
    \item $\mathit{replaceNode}(s1, s2)$: replace node $s1$ in $S1$ with node $s2$ in $S2$
    \item $\mathit{changeType}(s1, \textit{toType})$: change the type of node $s1$ to $\textit{toType}$
    \item $\mathit{changeChildrenOrder}(s1, \textit{toOrder})$: change the order of the children of node $s1$ to $\textit{toOrder}$
\end{itemize}

\subsection{Layout Tree Data Structure}

\begin{figure*}[t]
\centering
\includegraphics[width=\textwidth]{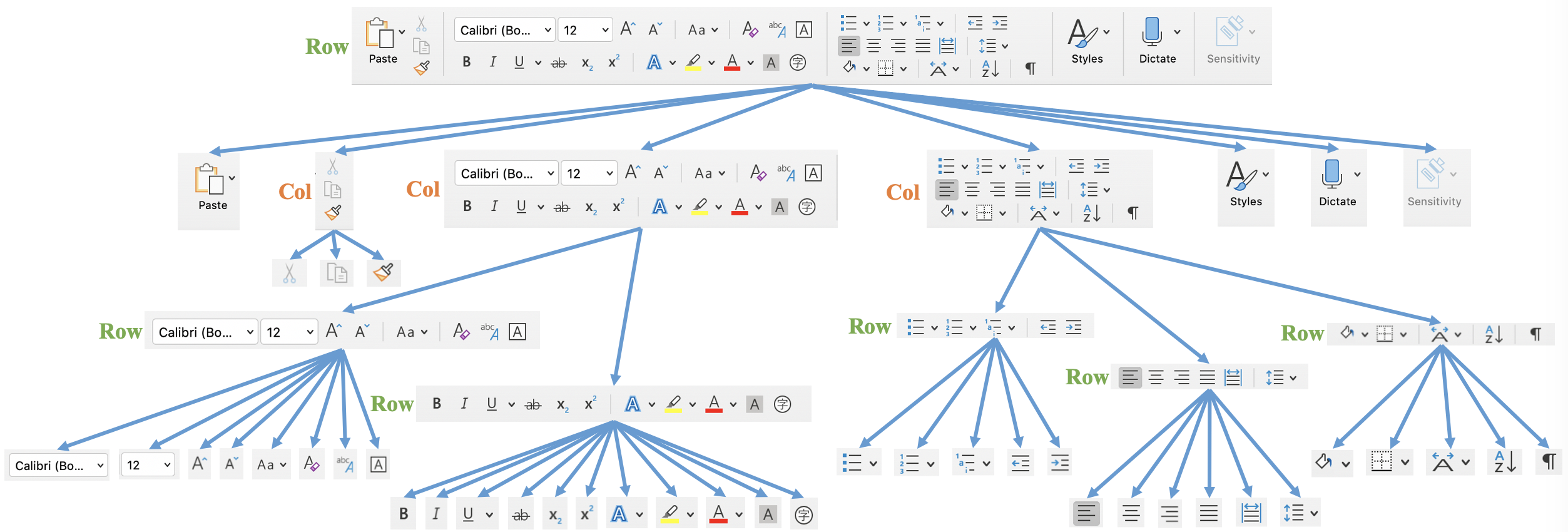}
\caption{Layout tree for the reconstructed layout structure of the MS Word ``ribbon'' in Figure \ref{fig:treeStructure}.}

\Description{This figure shows the reconstructed layout tree of the MS Word ``ribbon''. The root node of the tree is the original layout, The second level shows the sublayouts divided by vertical dividers and the third level shows the sublayouts further divided by horizontal dividers. Leaf nodes are all the buttons in the layout.}
\label{fig:tree}
\end{figure*}

We encode a reconstructed layout structure in a corresponding tree. Each widget becomes a leaf node, while each {\it Row} or {\it Column} becomes an internal node. For example, Figure \ref{fig:tree} shows the layout tree for the reconstructed layout structure of the MS Word ``ribbon'' in Figure \ref{fig:treeStructure}. Each node stores its properties: $widgetId$, $type$, $parent$, $children$, $pathToRoot$, $hashCode$, and $childHashCode$ (detailed descriptions of node properties are given in Appendix \ref{nodeProperties}). $hashCode$ is defined recursively: for leaf nodes is it $hash(widgetId)$ based on a standard hash function, and for internal nodes it is $hash(type) + 1 \times child1.hashCode + 2 \times child2.hashCode + 3 \times child3.hashCode + ...$, where $child1, child2, child3, etc.$ are children nodes of the current node. Similarly, $childHashCode$ is defined the same as $hashCode$ for leaf nodes, and as $child1.hashCode\ \textit{XOR} \ \ child2.hashCode\ \textit{XOR}$ $\ \ child3.hashCode\ \textit{XOR}\ \ ...$ for internal nodes.
$hashCode$ depends on the widget identity for leaf nodes, and the structure type, children nodes and their order for internal nodes. Thus, if two nodes in two layout specifications have the same $hashCode$, then they are identical with very high likelihood. $childHashCode$ only depends on the list of children of an internal node. It can be used to find corresponding nodes even if the node type ({\em e.g.}, changing from {\it Row} to {\it Column}) or the order of children has changed. 
We define two hash tables to keep track of $hashCode$ and $childHashCode$:
$hashMap$ maps the $hashCode$ of a node to the node itself, and $childHashMap$ maps from the $childHashCode$ of a node to the node itself.
When we compare two layout specifications, these two hash tables are used for quickly identifying corresponding nodes in both specifications (See more details in Appendix~\ref{nodeProperties}).

\subsection{Difference Detection Algorithm}

\begin{add}
We detect the differences between two layout specification trees
($\mathit{Tree1}$ and $\mathit{Tree2}$) and identify edit operations by recursively comparing corresponding lists of sibling nodes $S1$ and $S2$, with the corresponding lists containing child nodes of nodes that have already been determined to correspond. The basic idea is: we try to match nodes in $S2$ with corresponding nodes in $\mathit{Tree1}$. Whenever we have found a correspondence, we compare the respective nodes and record edit operations for any differences in position, type, or child order. 
We identify the corresponding nodes in the sibling lists and recursively apply this algorithm to the child nodes of corresponding nodes.


We first try to detect strong correspondences based on hash values to find all the nodes $s2 \in S2$ such that there is a node $s1 \in \mathit{Tree1}$ with the same $hashCode$ or $childHashCode$ as $s2$.
We then identify edit operations based on the differences between the corresponding nodes (for details see Appendix~\ref{identifyCorrespondNodes}). We expect most $s2 \in S2$ to have a corresponding node $s1 \in \mathit{Tree1}$. Thus, after this step, there should only be very few remaining nodes.
If we cannot find corresponding nodes for all the nodes $s2 \in S2$, we pair the remaining nodes in both trees based on their similarity. We keep pairing the remaining nodes depending on the largest number of common leaves and recursively call the algorithm on their child nodes (see Appendix~\ref{differenceDetection} for layout difference detection details). For example, in Figure~\ref{fig:overview} c, we identify move behaviors after detecting corresponding nodes in the two layout trees and comparing the node position differences. 


As most nodes typically can be easily paired based on their $hashCode$ and 
$childHashCode$, ReverseORC takes roughly linear time to process such nodes. Very few remaining nodes need to be paired based on similarities. Thus, the overall complexity of the layout difference detection algorithm is linear in practice for all GUI layouts we have tested.
\end{add}

\section{ORC Layout Specification Generation}

Based on the layout differences, ReverseORC infers layout behaviors and constructs corresponding ORC layout specifications, enabling later modification and customization. We use a pattern matching approach to find ORC Layout patterns that can describe the detected edit operations. Because layouts generally change incrementally, the set of detected edit operations usually only contains a small number of edit operations. These edit operations indicate the smallest changes in the UI and thus have a clear mapping to layout patterns, which enables us to perform precise pattern matching. 

\subsection{ORC Layout Pattern Matching}

ORC Layout~\cite{jiang2019orclayout} is one of the most flexible layout specification mechanisms that does not involve writing code. It comes with a set of layout patterns that can be used to specify common -- and also several not so common -- layout behaviours. By iterating over the layout transitions that we identified earlier on in the grid search, and considering each of the change sets identified by the tree difference detection, we identify and record ORC Layout patterns that can elicit the observed changes.
We start with the specification of the UI at its maximum size, and iterate `inwards' (right and up) over the samples and their change sets towards a UI's minimum size, i.e., in a way that describes a gradual change from the maximum to the minimum layout. For example, in the grid from Figure~\ref{fig:sampling}, change sets are considered in the following order: (Max) to (4) to (1), (Max) to (5) to (3) to (6), (3) to (8), (5) to (2) to (7) to (Min). 

In each iteration, we match a layout pattern to the respective change set and generalise the layout specification to include the respective pattern. The mapping between the edit operations in the change sets and the patterns is fairly direct, so patterns can be found by iterating over the edit operations and testing each pattern for applicability in a rule-based manner. If a pattern is applicable, we adjust our ORC layout specification to include the respective pattern. The most common edit operations and their associated patterns are as follows:

{\it removeNode(s1)}: If $s1$ is a leaf node, then $s1$ is an optional widget and we change the specification to mark it as such (``$s1$ is either there OR not''), using the current layout area (width $\times$ height) as penalty. As a result, a widget that disappears only when the layout gets small will have a small penalty, and the layout solver will implement this expected behavior. If $s1$ is a non-leaf node, this could be a knock-on effect of a flow layout with a {\it Row} or {\it Column} disappearing. We therefore check whether all the children of $s1$ have been moved away with corresponding {\it moveNode} operations. If that is not the case, $s2$ is marked as an optional sublayout. Otherwise, we ignore this operation as it will be handled by a different rule. 
{\it addNode(s2)}: This is the inverse case to {\it removeNode(s1)} and is handled analogously. If $s2$ is a non-leaf node, this could be a knock-on effect of a flow layout with a new {\it Row} or {\it Column} appearing, so we test this first.
{\it moveNode(s1, s2)}: If one or more consecutive nodes at the end of one Row / Column are moved to a (possibly new) adjacent Row / Column, then we merge Rows / Columns into a Horizontal / Vertical Flow. Otherwise, $s1$ has an alternative position at the location where $s2$ is, and we specify this by using ORC Layout's alternative position pattern (``$s1$ is either at position 1 OR at position 2'').
{\it replaceNode(s1, s2)}: $s1$ and $s2$ are alternative nodes, so we use ORC Layout's alternative layout pattern (``there is either $s1$ OR $s2$ at that location''). 
{\it changeType(s1, toType)}: $s1$ is marked as a pivot sublayout as a Row has changed to a Column or vice versa (``$s1$ is either a {\it Row} OR a {\it Column}'').
{\it changeChildOrder(s1, toOrder)}: $s1$ has an alternative widget order (``the children of $s1$ are either {\it fromOrder} OR {\it toOrder}'').

After we have detected a layout pattern, we use the API provided by ORC Layout~\cite{jiang2019orclayout, jiang2020orcsolver} to adjust the layout specification by instantiating and adding the observed pattern. 
If no patterns can be matched anymore and unmatched edit operations are still remaining, then this means that larger parts of the layout structure have simply been replaced by different layouts, e.g.\ as shown in Figure~\ref{fig:faultLines}. In this case we find the smallest subtree containing the respective changes and specify the two alternatives as logical disjunction (``either {\it subTree1} OR {\it subTree2}''. This is a sign of uncommon or drastic changes in the UI, as discussed below.

For example, in Figure~\ref{fig:teaser}, the difference between the first two layouts is that the ``Font'' button is replaced by its expanded version. According to the above rules, the edit operation $\mathit{replaceNode}$ indicates that this is an alternative layout pattern. The ``Font'' button and its expanded version are alternatives. The difference between the third and the forth layouts is that the ``Styles Pane'' button is added. As it is a widget (leaf node), the edit operation $\mathit{addNode}$ is mapped to an optional widget pattern.

\subsection{Visualizing Reconstructed Layout Quality}

\begin{figure*}[t]
\centering
\includegraphics[width=0.5\textwidth]{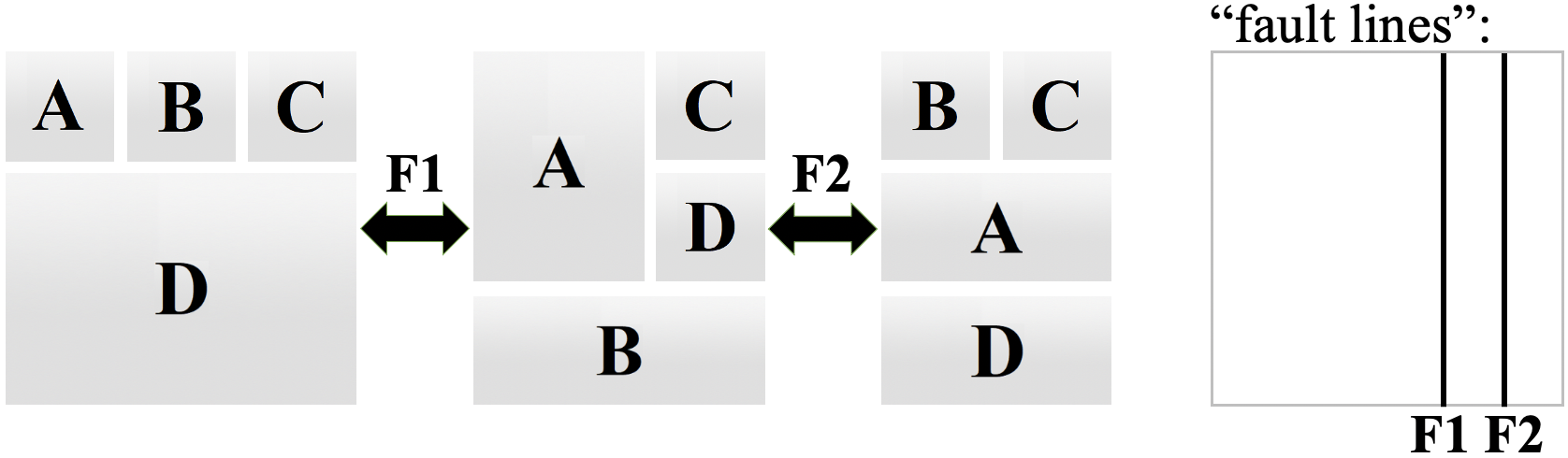}
\caption{Visualization of ``fault lines'' in the error map of a layout with `bad' (i.e.\ potentially confusing) behaviors.}

\Description{This figure shows a problematic layout and the corresponding error map. When the window is resized, the widgets are reordered in surprising ways. In the error map, the fault lines are shown as black lines indicating the transition positions where widgets are reordered problematically.}
\label{fig:faultLines}
\end{figure*}

\begin{figure*}[t]
\centering
\includegraphics[width=0.7\textwidth]{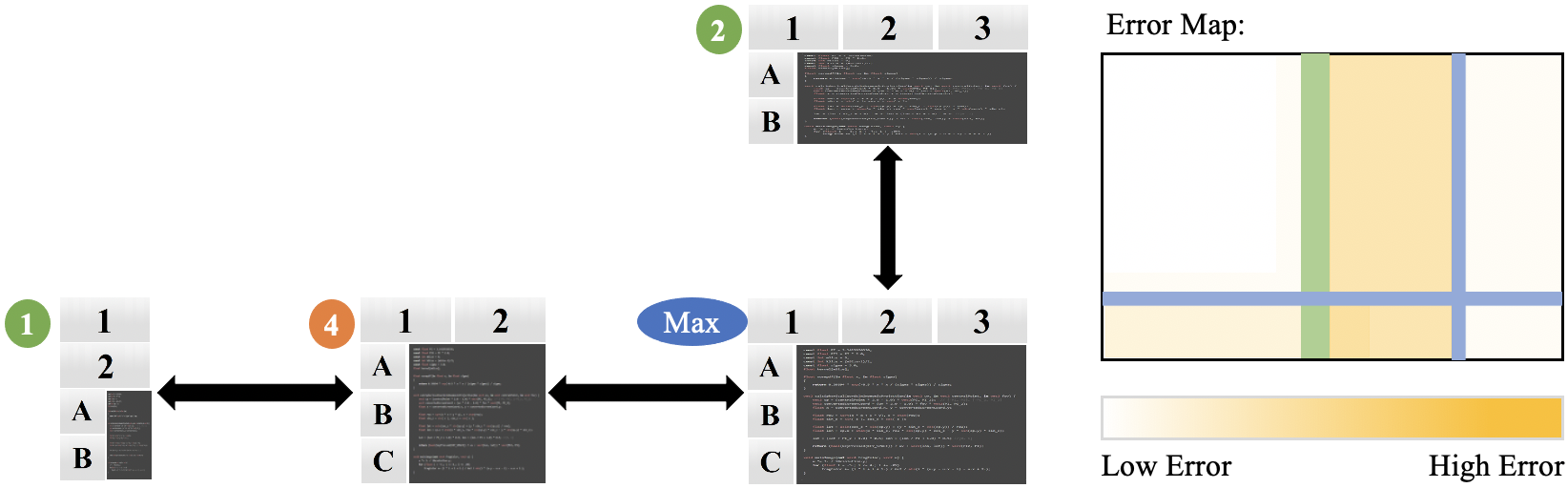}
\caption{\ADD{Visualization of reconstruction quality for the example in Figure \ref{fig:sampling} with an {\it error map}.}}

\Description{This figure shows some exemplars in Figure 3 with a error map corresponding to the subsampling example in Figure 3. The error map contains structural errors (darker yellow designates larger errors) and transition errors (blue / green), which measures the pixel difference between the transition positions of the layout specifications in the original and the reconstructed layout: green parts shows where the original GUI switches at a larger size than the reconstructed GUI, and blue for changes at a smaller size.}
\label{fig:error_map}
\end{figure*}

\ADD{To detect potential resize issues in layouts, designers often need a manual process to inspect the huge space of all potential device and layout dimensions to verify that there are no problems in the layouts. To address this challenge, we propose an error map that uses colors to help the designer pinpoint various interface dimensions that may be in need of improvement and/or repair. The map enables designers to see a visual overview of specific points in the resize space to enable them to quickly target and repair potential areas of concern. }

We visualize the quality of the reconstructed layout in an {\it error map} using three metrics: structural error, transition error, and ``fault lines''. The size of the error map matches the (scaled) size of the maximum layout, and sampled layout sizes correspond to points in the map. To define the structural error of a layout at a certain size, we consider the corresponding tabstops of the original layout and the corresponding reconstructed layout. The structural error of a layout is the sum of the squared differences between the positions of corresponding tabstops, divided by the number of tabstops. As illustrated in Figures~\ref{fig:teaser} and~\ref{fig:error_map}, the color of the error map at each point corresponds to the structural error of the sampled layout, with darker shades of yellow indicating larger error and linear color gradients filled in between the sampled points.

\begin{add}

As shown in Figure \ref{fig:error_map}, beyond structural error, the error map also visualizes the transition error (green / blue), which measures the pixel difference between the sizes of the original and the reconstructed layout at which a certain transition takes place ({\em e.g.}, a widget moving onto a new row): green parts indicate that the original GUI transitions at a larger size than the reconstructed GUI, and blue parts indicate that the original GUI transitions at a smaller size. For example, in the error map in Figure~\ref{fig:error_map}, the vertical blue area shows the transition error between (Max) and (4). The left boundary of the area is the transition position in the original layout and the right one is the transition position in the reconstructed layout. This area indicates that widget 3 disappears at a (slightly) smaller width in the original layout than the reconstructed layout. The vertical green area demonstrates the transition error between (4) and (1) indicating that widget 2 reflows to the next row at a (slightly) larger width in the original layout compared to the reconstructed layout. Analogously, the horizontal blue area shows that widget B disappears at a (slightly) smaller height in the original layout between (Max) and (2).

To highlight layout transitions that are potentially confusing to the user, {\em i.e.}, where widgets switch positions in surprising ways, we also identify such ``fault'' lines. In Figure \ref{fig:faultLines} we show such a problematic layout, where the fault lines F1 and F2, which correspond to the transitions on the left, are shown as black lines in the error map on the right. More specifically, we show fault lines when a) widgets are reordered or b) larger parts of the layout structure change, as indicated by layout alternatives that cannot be matched to common ORC Layout patterns, {\em i.e.}, where an OR needs to be inserted between two whole sublayout alternatives. Fault lines indicate transition positions that might need adjustment in the reverse engineered specification.  
In Figure~\ref{fig:faultLines}, we show a layout with resize behaviors that cannot be predicted with common ORC Patterns. Both transition positions have fault lines, which illustrate the points where an unpredictable behavior occurs and where the layout could potentially be improved. Thus, designers could use fault lines as guides to identify and fix bad layout behaviours, {\em e.g.}, by modifying the generated ORC layout specification. For example, this layout could be changed to a horizontal flow layout to exhibit better resize behavior.
\end{add}

\section{Applications}

It is often time-consuming for designers to create new resizable UIs from scratch. Sometimes a designer might find a UI that is similar to what they are looking for. ReverseORC can help designers to reconstruct ORC Layout specifications for existing UIs and then use those specifications in other applications and on other platforms. In the following, we briefly discuss this for the MS Word ``ribbon'' GUI and the BBC News website, which both use highly dynamic layouts.
Furthermore, we briefly discuss how ReverseORC can help designers to modify, extend and even create resizable GUIs by example.

\subsection{GUI Reverse Engineering -- MS Word Ribbon}

ReverseORC can be used on dynamic, hand-coded GUI layouts, such as the well-known MS Word ``ribbon'' toolbar. In Figure \ref{fig:teaser}, we present our reverse engineering result for the ``ribbon''. The yellow lines in the original UI samples on the left illustrate the layout structure results of each sample. The edit operations detected for the transitions between them are shown with blue arrows. On the right, the corresponding reconstructed UI with its ORC Layout patterns are shown, exhibiting the same layout behaviours as the original. 


\begin{figure*}[t]
\centering
\includegraphics[width=0.85\textwidth]{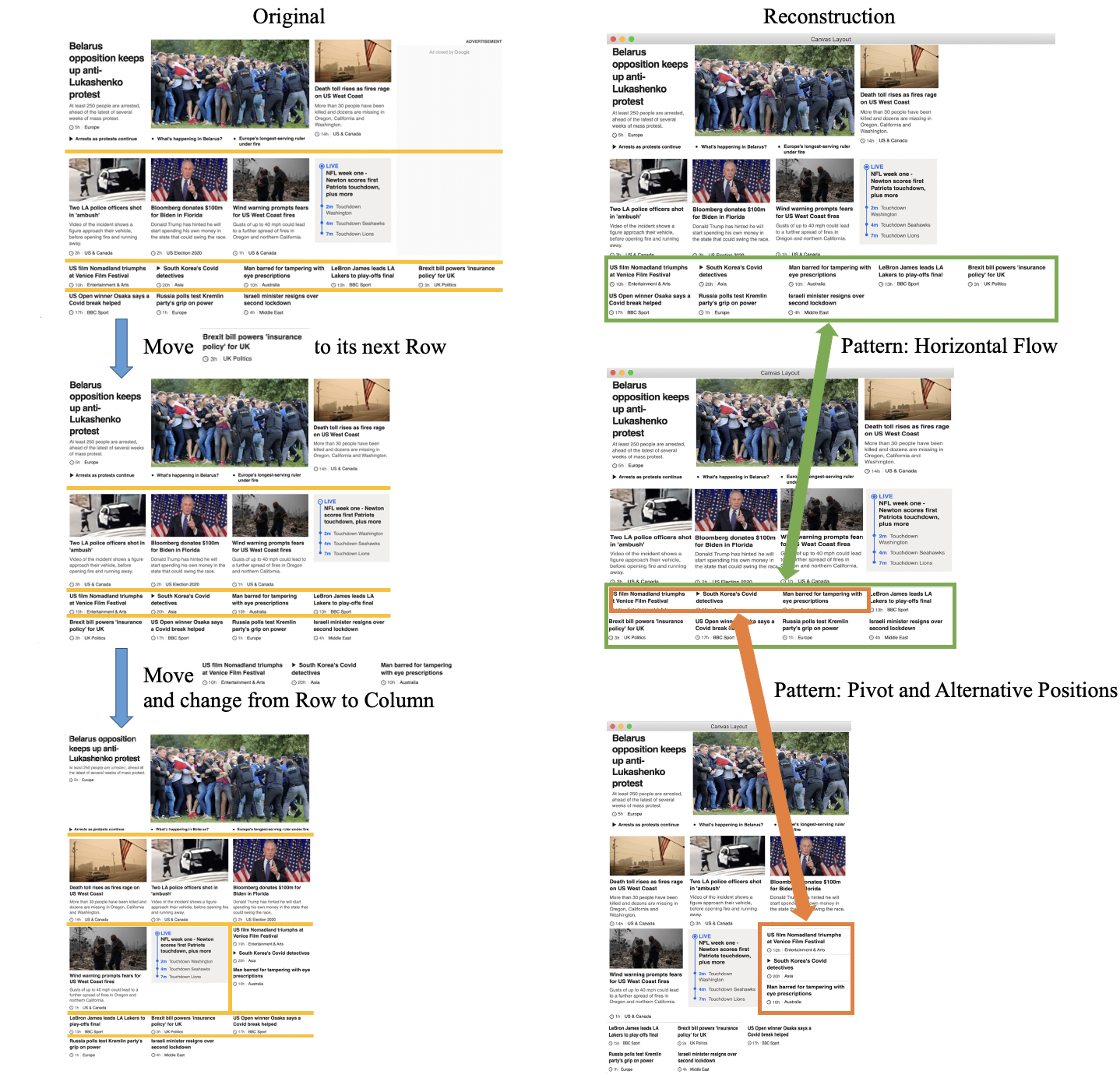}
\caption{Reverse engineering result for the BBC News website, displayed as an ORC Layout GUI. One Horizontal Flow pattern was omitted in the figure for space reasons.}

\Description{This figure shows the appearance of original the BBC News website and the reconstructed BBC News website using ReverseORC at three different sizes. Horizontal lines on the original website are divisors that define Row layout containers and vertical ones define Columns. We show the changes between neighbouring sizes of the original website and the corresponding layout patterns between neighbouring sizes of the reconstructed website.}
\label{fig:bbc_news}
\end{figure*}

\subsection{Web UI Reverse Engineering -- BBC News}

ReverseORC enables moving layouts across platforms. Designers may want to replicate a web layout in a mobile app or vice versa. As our system is platform and framework independent, this means that a layout can be re-used in another form of applications, as we can unify different layouts by reverse engineering. For example, we can reverse engineer GUI layouts for the web, and web layouts for GUIs. As a demonstration of this, Figure~\ref{fig:bbc_news} shows the reverse engineering result for the BBC News website into a GUI environment, which opens up options for cross-platform applications.  Our method works well for webpages that have well-defined and reasonably predictable layout methods, but we cannot claim that our method works well for all layout methods that exist on the web. Consider for example a tiled layout that rearranges tiles randomly upon a resize. In this case, ReverseORC creates a very large layout specification that contains many OR clauses.

\begin{figure*}[t]
\centering
\includegraphics[width=0.5\textwidth]{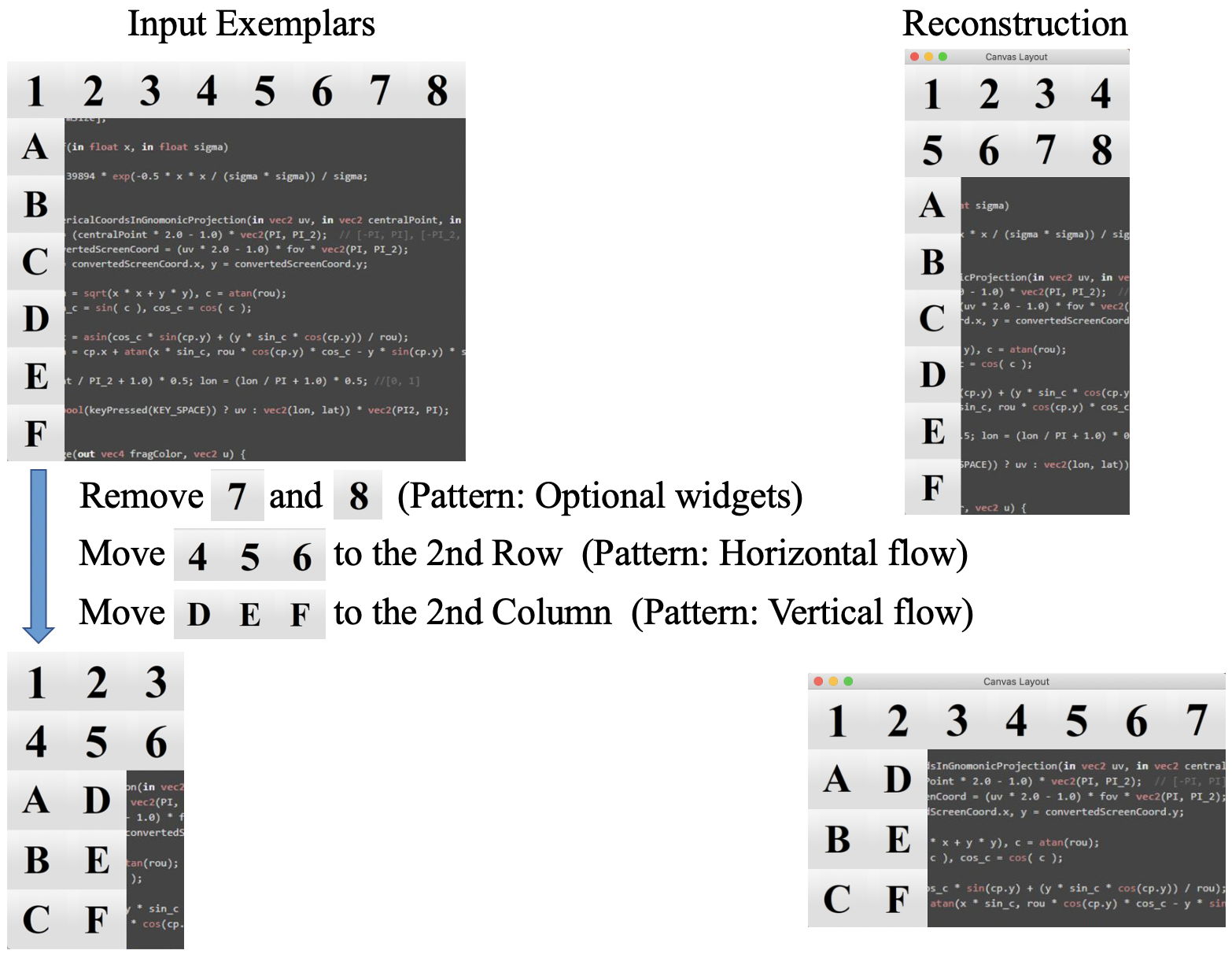}
\caption{Given multiple different sizes of a static GUI layout drawn by a visual designer, ReverseORC can generate an ORC layout specification based on the patterns inferred from the drawn layout results. The two figures on the right show the results of resizing the GUI reconstructed from the left two window sizes.}

\Description{The left portion of the figure shows two different sizes of GUI layout drawn by a visual designer. Changes detected by ReverseORC are shown between the two layouts. The left portion of the figure shows the reconstructed layout.}
\label{fig:application}
\end{figure*}

\subsection{Exemplar-Based Layout Design}
\label{examplarbasedLayoutDesign}

The layout difference detection and ORC layout pattern matching parts of ReverseORC can be used to reverse engineer desired layout behaviors and generate resizable GUIs based on examples, {\em e.g.}, multiple different sizes of a static GUI layout drawn by a visual designer. Given multiple such static GUI layouts drawn by a visual designer for different window sizes, ReverseORC detects layout differences among the static exemplar layouts and infers an ORC layout that matches the drawn layout results and the designer's intention (Figure~\ref{fig:application}). 

If the results do not match the designer's expectations, they can iteratively draw new exemplars, or change their existing exemplars, and ReverseORC's difference detection will pick up the differences and change the layout specification accordingly. This can be used, for example, to disambiguate some layout behaviors by providing more examples, or add extra transitions for a smoother resize behaviour. The designer could pick a respective size by clicking on the error map, and then modify or replace the UI for the chosen size in a UI builder-like interface.

Similarly, manual exemplars can be combined with exemplars that are sampled automatically. For example, in Figure~\ref{fig:teaser}, the transition at the fault line between layout four and five rearranges two sublayouts in a fairly arbitrary manner to use available width. Upon seeing the fault line, a designer could manually re-draw layout five, e.g.\ in a manner that re-arranges the widgets according to a flow layout. ReverseORC would then create a Flow pattern for the transition and the fault line would disappear.



\section{Discussion}

ReverseORC is an implementation- and platform-independent reverse engineering approach to detect layout behaviors and types and to generate a matching high-level ORC Layout specification for a given layout. It enables the creation of responsive, flexible layouts for existing and new applications. ReverseORC is an efficient tool; it took only about 0.4 seconds to reverse engineer the MS word ``Ribbon'' toolbar and about 0.5 second for the BBC News website on an average laptop computer. It can be used as part of the design process for making existing UIs more flexible, to fix problems in existing GUIs, and in developing completely new GUIs. 
In effect, we can deal with most layout methods, including flow, grid, grid-bag, and ORC layouts. One caveat is that we currently cannot reconstruct some of the numerical parameters influencing a layout, such as the weights in grid-bag layouts and their ability to center content. This is a topic for future work.


\begin{add}
Developers interact with ReverseORC by editing UI exemplars (Section \ref{examplarbasedLayoutDesign}). Such editing of a UI’s appearance has been well studied~\cite{scoditti2009new, zeidler2012auckland} and  found to be developer-friendly, especially when compared to specifying interactive behaviors directly~\cite{myers2008designers}. Similar to Expresso~\cite{krosnick2018expresso}, ReverseORC developers also specify UI exemplars simply by dragging and resizing elements, and this was already demonstrated to be easy and useful by Expresso. The usefulness of the resulting ORC specifications has been validated in \cite{jiang2019orclayout, jiang2020orcsolver}.
\end{add}

A drawback of using accessibility APIs to extract widget information is that in some applications, not all the widgets might provide accessibility APIs. Hybrid techniques combining pixel-based methods and accessibility API could further improve the accuracy of widget detection~\cite{hurst2010automatically}. 
For websites, sampling is predominantly a one-dimensional problem as window widths are much more important. Due to the affordance of vertical scrolling, heights are relatively less relevant. Thus, instead of a binary grid search, a simpler approach for website layouts might be to sample different widths through a one-dimensional binary interval search.

If the designer or the implementer of a layout manager made a severe mistake when a GUI was designed, which causes unexpected widget placement to occur in a layout, e.g., an optional widget that ``flickers in and out'' during resizing, then our framework will typically create many alternative patterns -- since our ReverseORC approach can only detect known layout types and patterns. While this is a fundamental limitation of our approach, it is not an algorithmic one, as we are in this case not dealing with well-defined layout behaviors. On other other hand, our approach could also be used as a sanity test for layouts to detect bugs and/or unexpected behaviors, as fault lines would appear in the error map for many such behaviors. The user can then use the results of our algorithm to replace unexpected behaviors in the layout with more deterministic and predictable patterns.
\section{Conclusion and Future Work} 

We presented ReverseORC, a novel layout reverse engineering method that reconstructs layout specifications for existing UIs, considering not only the static structure of the original but also its dynamic resize behaviors. By sampling layout sizes with a binary grid search, ReverseORC detects topological differences between layouts of different sizes, further infers layout behaviors, and generates a corresponding ORC Layout specification to enable layout customization and generation of new UIs. To our knowledge, ReverseORC is the first approach for reverse engineering dynamic resizable layouts and generating a platform independent high-level layout specification for them. We envision that our method could be widely applied in various applications and platforms. ReverseORC is available as open source from \url{https://github.com/YueJiang-nj/ReverseORC-CHI2021}.

\bibliographystyle{ACM-Reference-Format}
\bibliography{Reference}
\appendix

\section{Tabstop Creation}
\label{tabstopCreation}

 \begin{algorithm}[t]
    \SetKwInOut{Input}{Input}
    \SetKwInOut{Output}{Output}
\SetAlgoLined

\SetKwProg{Fn}{Function}{}{end}
\Fn{\textit{TabstopCreation}(L)}{
    
    $xtabs: (\mathbb{Z} \rightarrow \textit{X-Tabstop}) = \{0 \mapsto L.\textit{leftTab},\ L.width \mapsto L.rightTab\}$
    
    $ytabs: (\mathbb{Z} \rightarrow \textit{Y-Tabstop}) =  \{0 \mapsto L.\textit{topTab},\ L.height \mapsto L.bottomTab\}$
    
    \For{$w \in L.\textit{Widgets}$} {
    
        \If  {$w.\textit{left} \in Domain(xtabs$)} {
            $w.\textit{leftTab} \leftarrow xtabs(w.\textit{left})$
        } \Else {
            $xtabs \leftarrow \textit{xtabs}\ \cup\ \{w.\textit{left} \mapsto w.\textit{leftTab}\}$
        }
        
        Process right, top, bottom tabstops analogously.
        
        
        
    }
    \Return $xtabs$, $ytabs$
}

\caption{Tabstop Creation}
\label{alg:tabstopcreation}
\end{algorithm}

For each layout $L$, we define tabstops through two functions $xtabs()$ and $ytabs()$ that map from positions in the GUI to tabstop variables in the layout (Algorithm \ref{alg:tabstopcreation}). We define $xtabs$ as a function mapping from x-coordinates to x-tabstops. Initially, it contains two elements that map the leftmost x-position in the GUI to the left tabstop variable of the layout and correspondingly for the right (Line 2). $ytabs$ is the analogous mapping for y-coordinates to y-tabstops (Line 3). 

Line 4-12 show how we loop through all the widgets in the GUI to create mappings to the two tabstop functions $xtabs$ and $ytabs$. We check if the x-coordinate of the current widget's left boundary $w.\textit{left}$ is contained in the domain of xtabs, {\em i.e.}, this x-coordinate $w.\textit{left}$ is already mapped to an existing x-tabstop in the function $xtabs$ (Line 5). In practice, there might be some small displacements in the layout. For example, a widget might be displaced by a pixel due to a rounding error. Then it is unreasonable to add two tabstops with a one-pixel distance in between. Thus, instead of checking whether $w.\textit{left}$ is in the domain of the function $xtabs$, we can check whether there exists an x-coordinate $xpos$ in $xtabs$ that is within a tolerance value $\epsilon$. If so, we map $w.\textit{leftTab}$ to $xtabs(w.left)$ to eliminate near-duplicate tabstop variables and near-identical mappings in the function $xtabs$ (Line 6). If $w.\textit{left}$ could not be mapped to a tabstop variable in the function $xtabs$, then we insert a new mapping from the x-coordinate $w.\textit{left}$ to the tabstop variable $w.\textit{leftTab}$ (Line 9). We process all three other boundaries of each widget (right, top, bottom) analogously.

In the end, the algorithm yields the final $xtabs$ and $ytabs$ functions. Also, we now have four unique tabstop variables for each widget in the layout.

\section{Tabstop Layout Dividers}
\label{tabstopLayoutDivider}

 \begin{algorithm}[t]
    \SetKwInOut{Input}{Input}
    \SetKwInOut{Output}{Output}
\SetAlgoLined


\SetKwProg{Fn}{Function}{}{end}
\Fn{\textit{\textit{GetTabstopLayoutDividers}}(tabs, widgets, xy)}{
    
    $\textit{tabValues} \leftarrow sorted(tabs.keys())[1:-1]$
    
    $\textit{tabstopLayoutDividers} \leftarrow []$
    
    \For{$tabValue \in tabValues$} {
    
        $divideLayout \leftarrow True$
        
        \For {$w \in widgets$} {
            \If {$xy == "x"$}{
                $minBoundary \leftarrow w.left$
                
                $maxBoundary \leftarrow w.right$
            } 
            \If {$xy == "y"$}{
                $minBoundary \leftarrow w.top$
                
                $maxBoundary \leftarrow w.bottom$
            } 
            \If {$minBoundary < tabValue$ and $maxBoundary > tabValue$} {
                $divideLayout \leftarrow False$
            }
        }
        \If {$divideLayout == True$} {
            $\textit{tabstopLayoutDividers} \leftarrow \textit{tabstopLayoutDividers} \cup \{tabValue\}$
        }
    }
    \Return $\textit{tabstopLayoutDividers}$
}
\caption{Tabstop Layout Dividers}
\label{alg:tabstoplayoutdividers}
\end{algorithm}

We define a tabstop as a tabstop layout divider if it is a clean cut dividing the layout into two parts where the tabstop does not cross any widget in the layout. 
For a horizontal tabstop layout divider, all the widgets in the layout are either above it or below it, analogously for vertical tabstop layout dividers. We first get all the tabstops that divide the layout into two parts (Line 2). Then we loop over each of these tabstops, we check if all the widgets in the layout have minimum boundary greater than the tabstop value or maximum boundary less than the tabstop. (For the x-axis, the minimum boundary of a widget is its left boundary and the maximum its right boundary, while for y-axis, they are the top and bottom boundaries respectively.) If so, the tabstop is a clean cut for the layout, and thus a tabstop layout divider (Line 4-22). 

\section{Layout Structure Reconstruction}
\label{layooutStructureConstruction}

 \begin{algorithm}[t]
    \SetKwInOut{Input}{Input}
    \SetKwInOut{Output}{Output}
\SetAlgoLined
\SetKwProg{Fn}{Function}{}{end}
\Fn{\textit{ConstructLayoutStructure}(L)}{
    if layout is a single widget, return it \\
    $xtab, ytab \leftarrow \textit{TabstopCreation(L)}$
    
    $ytabLayoutDividers \leftarrow \textit{GetTabstopLayoutDividers(ytab, L.widgets, "y")}$
    
    $layoutStructure \leftarrow \{"Column" : []\}$
    
    \If {$ytabLayoutDividers$ not empty} {
        
        $widgetList \leftarrow []$
    
        $widgetsCurr \leftarrow L.widgets \text{~sorted by~} w.bottom$
        
        \For {$\textit{tabValue} \in ytabLayoutDividers$} {
            $\textit{sublayoutWidgets} \leftarrow \{ w\ | w \in widgetsCurr \wedge w.bottom \leq \textit{tabValue} \}$
            
            if $\textit{sublayoutWidgets}$ is empty, then continue
            
            $layoutStructure["Column"] \leftarrow \quad layoutStructure["Column"] \cup \{ConstructLayoutStructure(Layout(\newline
            \textit{sublayoutWidgets}))\}$
            
            $widgetList \leftarrow widgetList \cup \{\textit{sublayoutWidgets}\}$
            
            $widgetsCurr \leftarrow widgetsCurr - \textit{sublayoutWidgets}$ 
        }
        \If {$widgetsCurr\ not\ empty$} {
            $layoutStructure["Column"] \leftarrow layoutStructure["Column"] \cup \{ConstructLayoutStructure(Layout(\newline
            \textit{sublayoutWidgets}))\}$
            
            $widgetList \leftarrow widgetList \cup \{\textit{widgetsCurr}\}$
        }
        
        $layoutStructure["Column"][i:j] \leftarrow ConstructLayoutStructure(Layout(\newline
        merge(widgetList[i:j])))$ if possible simplified structure exists for any $0 \leq i < j < len(layoutStructure["Column"])$
        
            
                
        \Return $layoutStructure$
    }
    process $xtab$ analogously
    
    if no subdivision is possible, then return $L$
}
\caption{Layout Structure Construction}
\label{alg:layouttreeconstruction}
\end{algorithm}

We reconstruct the layout structure by recursively subdividing it based on layout dividers (Algorithm~\ref{alg:layouttreeconstruction}). As the basic case in this recursion, if the current sublayout only contains a single widget, we simply return its identity (Line 1). Otherwise, we first try to subdivide the layout using vertical layout dividers (Line 3-4). If such subdivision is possible (Line 6), then we sort all the widgets in the layout by their bottom boundary positions (Line 8). We then assign the widgets to different sublayouts based on the positions of the horizontal layout dividers, and recursively use the reconstruction on each sublayout structure (Line 9-19). We merge two layout dividers into one if there is no widgets between them (Line 11). 

We aim to reconstruct the simplest possible layout structure. Therefore, to avoid creating layout dividers caused by accidental alignments, we regroup widgets in multiple consecutive sublayouts and try running the algorithm recursively to simplify the resulting layout structure. We reconstruct the sublayout if we can get a simplified sublayout structure by grouping them (Line 20).
We try horizontal subdivision (with vertical layout dividers) first as it is more common and in line with reading order. If horizontal subdivision is not possible, we process vertical subdivision analogously (Line 23). If both cases are impossible, which is very rare as UIs are typically laid out using a division-based containment hierarchy, then the layout can only be described using tabstops directly (Line 24), {\em e.g.}, in a pinwheel layout \cite{zeidler2017tiling}. Figure \ref{fig:treeStructure} shows the visualization of the constructed layout structure of the MS Word ``ribbon''.

\section{Node Properties}
\label{nodeProperties}

Based on the resulting layout tree structure of an input layout specification gotten from Algorithm \ref{alg:layouttreeconstruction}, we traverse this tree structure and create a corresponding new tree. Each widget becomes a leaf node, while each $Row$ or $Column$ structure becomes an internal node (non-leaf node). Each node stores the following properties:
\begin{itemize}
    \item $widgetId\ /\ type$: 
        \subitem - leaf nodes: widget identifier 
        \subitem - internal nodes: structure type (either $"Row"$ or $"Column"$)  
    \item $parent$: parent node of the current node
    \item $children$: the list of children nodes of the current node
    \item $pathToRoot$: a list of tuples containing the ancestors of the current node along with the positions among their siblings, ({\em e.g.,} If $pathToRoot$ of the current node is $[(Root, 2), (A, 3), \newline (B, 4)]$, then $A$ is the 2nd child of the $Root$ node, $B$ is the 3rd child of $A$, and the current node is the 4th child of $B$.)
    \item $leaves$: the list of all the leaves in the subtree rooted at the current node
    \item $hashCode$: 
        \subitem - leaf nodes: $hash(widgetId)$ based on a standard hash function 
        \subitem - internal nodes: $hash("Row"/"Column") \newline + 1 \times child1.hashCode + 2 \times child2.hashCode \newline+ 3 \times child3.hashCode + ...$, \newline where $child1, child2, child3, etc.$ are children nodes of the current node
    \item $childHashCode$:
         \subitem - leaf nodes: same as $hashCode$
        \subitem - internal nodes: $child1.hashCode\ \textit{XOR} \ \ child2.hashCode\ \newline \textit{XOR} \ \ child3.hashCode\ \textit{XOR}\ \ ...$, where $child1, child2, etc.$ are children nodes of the current node
\end{itemize}

\section{Identifying Corresponding Nodes}
\label{identifyCorrespondNodes}

 \begin{algorithm}[t]
    \SetKwInOut{Input}{Input}
    \SetKwInOut{Output}{Output}
\SetAlgoLined
\SetKwProg{Fn}{Function}{}{end}
\Fn{\textit{\textit{DetectIdenticalNode}}(S1Curr, S2Curr)}{
    \For {$s2 \in S2Curr$} {
        \If {$s2.hashCode \in S1HashMap.keys()$} {
            $S1Curr \leftarrow S1Curr - s1$ if $s1 \in S1$
            
            $S2Curr \leftarrow S2Curr - s2$
            
            \If {$s1 \notin S1$} {
                $changes$ += $\{moveNode(s1, s2)\}$
            } \Else {
                $pairs \leftarrow pairs \cup \{s2 \mapsto s1\}$
            }
        }
    }
}
\caption{Identical Node Detection}
\label{alg:identicalnodedetection}
\end{algorithm}

 \begin{algorithm}[t]
    \SetKwInOut{Input}{Input}
    \SetKwInOut{Output}{Output}
\SetAlgoLined
\SetKwProg{Fn}{Function}{}{end}
\Fn{\textit{\textit{DetectSimilarNodes}}(S1Curr, S2Curr)}{
    \For {$s2 \in S2Curr$} {
        \If {$s2.childHashCode \in S1ChildHashMap.keys()$} {
            $S1Curr \leftarrow S1Curr - s1$ if $s1 \in S1$
            
            $S2Curr \leftarrow S2Curr - s2$
            
            \If {$s1 \notin S1$} {
                $changes \leftarrow changes \cup \{moveNode(s1, s2)\}$
            } \Else {
                $pairs \leftarrow pairs \cup \{s2 \mapsto s1\}$
            }
            \If {$s1.type \neq s2.type$} {
                $changes \leftarrow changes \cup \{changeType(s1, s1.type, s2.type)\}$
            }
            \If {$s1.children \neq s2.children$} {
                $changes \leftarrow changes \cup \{changeChildrenOrder(s1, s1.children, s2.children)\}$
            }
        }
    }
}
\caption{Similar Node Detection}
\label{alg:similarnodedetection}
\end{algorithm}

To identify corresponding nodes in the sibling lists $S1$ and $S2$, we first loop over the $S2Curr$ list to find all the nodes $s2 \in S2$ such that there is a node $s1 \in Tree1$ with the same $hashCode$ as $s2$ (Algorithm \ref{alg:identicalnodedetection}). If the corresponding node $s1$ does not belong to $S1$, then $s1$ moved to $S2$ at the position it occurs in $S2$ (Line 6-8). Otherwise, we pair $a1$ and $s2$ in sibling lists (Line 10). We expect most $s2 \in S2$ have its corresponding node $s1 \in Tree1$. Thus, after this step, $S2Curr$ list should only contain very few nodes. Similarly, for all the remaining nodes $s2 \in S2Curr$, we check whether there is a node $s1 \in Tree1$ with same $childHashCode$ as $s2$, and pair them accordingly (Algorithm \ref{alg:similarnodedetection} Line 2-11). In addition, we identify type changes and children node order changes (Line 12-17).  

\section{Layout Difference Detection}
\label{differenceDetection}

 \begin{algorithm}[t]
    \SetKwInOut{Input}{Input}
    \SetKwInOut{Output}{Output}
\SetAlgoLined
\SetKwProg{Fn}{Function}{}{end}
\Fn{\textit{\textit{LayoutDifferenceDetection}}(S1, S2)}{
    $pairs \leftarrow \{\}$ // hash table from nodes in S2 to nodes in S1
    
    $S1Curr \leftarrow S1$
    
    $S2Curr \leftarrow S2$
    
    $\textit{DetectIdenticalNodes}(S1Curr, S2Curr)$ // find and pair S2 nodes that are identical to {\it Tree1} nodes
    
    $\textit{DetectSimilarNodes}(S1Curr, S2Curr)$ // find and pair S2 nodes that have the same leaves as {\it Tree1} nodes
    
    \While{$S2Curr$ not empty}{
        $(s1Best, s2Best) \leftarrow (s1\in S1, s2\in S2)$ {\em s.t.} max number of common leaves
        
        $maxSim \leftarrow numCommonLeaves(s1Best, s2Best)$
        
        \If {$maxSim > 0$} {
            $S1Curr \leftarrow S1Curr - s1$ 
            
            $S2Curr \leftarrow S2Curr - s2$
            
            $pairs \leftarrow pairs \cup \{s2Best \mapsto s1Best\}$
            
            \If {$s1Best.type \neq s2Best.type$} {
                $changes \leftarrow changes \cup \{changeType(s1Best, s2Best.type)\}$
            }
            $\textit{LayoutDifferenceDetection}(s1Best.children,\newline s2Best.children)$
        } \Else {
            break
        }
    } 
    $S1Paired \leftarrow [s1\ for\ s1 \in S1\ if\ s1\in pairs.values()]$
    
    $S2Paired \leftarrow [s2\ for\ s2 \in S2\ if\ s2\in pairs.keys()]$
    
    Replace $s2 \in S2Paired$ by $pairs[s2]$
    
    \If {$S1Paired \neq S2Paired$} {
        $changes \leftarrow changes \cup \{changeChildrenOrder(s1.parent, S2Paired)\}$
    }

    \If {num of paired nodes before $s1\in S1$ = num of paired nodes before $s2\in S2$} {
        $S1Curr \leftarrow S1Curr - s1$ 
            
        $S2Curr \leftarrow S2Curr - s2$
        
        $changes \leftarrow changes \cup \{replaceNode(s1, s2)\}$
    }
    
    \For {$s1 \in S1Curr$} {
        $changes \leftarrow changes \cup \{removeNode(s1)\}$
    }
    
    \For {$s2 \in S2Curr$} {
        $changes \leftarrow changes \cup \{addNode(s2)\}$
    }
}
\caption{Layout Difference Detection}
\label{alg:layoutdifferencedetection}
\end{algorithm}

We detect the differences between two layout specifications and identify edit operations by recursively comparing corresponding lists of sibling nodes $S1$ and $S2$ (Algorithm \ref{alg:layoutdifferencedetection}), with the corresponding lists containing child nodes of nodes that have already been determined to correspond. The basic idea is: we try to match corresponding nodes in S2 with nodes in Tree1. Whenever we have found a correspondence, we compare the respective nodes and record edit operations for any differences in position, type or child order. Initially, the inputs of the layout difference detection algorithm are $S1 = [root\ node\ \mathit{of}\ \mathit{Tree1}]$ and $S2 = [root\ node\ \mathit{of}\ \mathit{Tree2}]$, where $\mathit{Tree1}$ and $\mathit{Tree2}$ are the trees representing the two layout specifications. We then identify the corresponding nodes in the sibling lists and recursively apply this algorithm to the children nodes of corresponding nodes. As we find corresponding nodes, we specify their differences (if any) as edit operations and add them to a set {\it changes}. In each call, we maintain a hash table $pairs$ mapping nodes in $S2$ to corresponding nodes in $S1$; with $pairs$ initially empty (Line 2). We keep track of currently unpaired nodes by removing the paired nodes from the lists $S1$ and $S2$ once a pair has been found.

We first try to detect strong correspondences based on a node's hash values in {\it IdenticalNodeDetection} and {\it DetectSimilarNodes}. These methods loop over $S2Curr$ to find all the nodes $s2 \in S2$ such that there is a node $s1 \in \mathit{Tree1}$ with the same $hashCode$ or $childHashCode$ as $s2$. If the corresponding node $s1$ does not belong to $S1$, then $s1$ moved to $S2$ at the position it occurs in $S2$.  Otherwise, we pair $a1$ and $s2$ in sibling lists. In addition, we identify type changes and also children node order changes (Details see Appendix Section \ref{identifyCorrespondNodes}). We expect most $s2 \in S2$ have its corresponding node $s1 \in \mathit{Tree1}$. Thus, after this step, $S2Curr$ list should only contain very few nodes. 

If we cannot find corresponding nodes for all the nodes $s2 \in S2Curr$, we pair the remaining nodes in $S1Curr$ and $S2Curr$ based on their similarity. We keep pairing $(s1\in S1Curr, s2\in S2Curr)$ depending on the largest number of common leaves and recursively call the algorithm on their children nodes. We stop this pairing process when there is no node remaining in $S2Curr$ or all the $s1\in S1Curr$ and $s2\in S2Curr$ have no common leaves (Algorithm \ref{alg:layoutdifferencedetection} Line 7-22). In addition, we check whether the order of all the paired nodes has not changed in $S1$ and $S2$ (Line 23-28).

After all the above pairing operations, $S1Curr$ and $S2Curr$ contain all the nodes that cannot be paired with any node in the other layout tree. If $s1\in S1Curr$ and $s2\in S2Curr$ have the same relative position among their sibling nodes, {\em i.e.,} the number of paired nodes before them are the same, then we infer that $s1$ in $S1$ is replaced by $s2$ in $S2$ (Line 29-33). All the remaining $s1\in S1Curr$ are removed from $S1$ and $s2\in S2Curr$ are added in $S2$ (Line 34-39). 

\end{document}